\documentclass[reprint,prl,superscriptaddress,showpacs,twocolumn]{revtex4-1}
\usepackage{units}
\usepackage{amsmath}
\usepackage{amssymb}
\usepackage{graphicx}
\usepackage{bm}
\usepackage{microtype,color}

\newcommand{\be}{\begin{equation}}
\newcommand{\ee}{\end{equation}}
\newcommand{\bea}{\begin{eqnarray}}
\newcommand{\eea}{\end{eqnarray}}

\begin{document}

\title{Shape-dependence of transmission, reflection and absorption eigenvalue densities \\
in disordered waveguides with dissipation}

\author{A. Yamilov}
\email{yamilov@mst.edu}
\affiliation{Department of Physics, Missouri University of Science and Technology, Rolla, Missouri 65409,USA}
\author{S. Petrenko}
\affiliation{Department of Physics, Missouri University of Science and Technology, Rolla, Missouri 65409,USA}
\author{R. Sarma}
\affiliation{Department of Applied Physics, Yale University, New Haven, CT, 06520, USA}
\author{H. Cao}
\email{hui.cao@yale.edu}
\affiliation{Department of Applied Physics, Yale University, New Haven, CT, 06520, USA}

\date{\today}
\begin{abstract}
The universal bimodal distribution of transmission eigenvalues in lossless diffusive systems underpins such celebrated phenomena as universal conductance fluctuations, quantum shot noise in condensed matter physics and enhanced transmission in optics and acoustics.
Here, we show that in the presence of absorption, density of the transmission eigenvalues depends on the confinement geometry of scattering media. 
Furthermore, in an asymmetric waveguide, densities of the reflection and absorption eigenvalues also depend of the side from which the waves are incident. 
With increasing absorpotion, the density of absorption eigenvalues transforms from single-peak to double-peak function.
Our findings open a new avenue for coherent control of wave transmission, reflection and absorption in random media.
\end{abstract}
\pacs{42.25.Dd, 42.25.Bs, 72.15.-v}
\maketitle
%42.25.Bs Wave propagation, transmission and absorption
%42.25.Dd Wave propagation in random media
%42.25.Fx Light scattering: wave optics
%11.55.-m Scattering matrix
%72.15.-v Electronic conduction in metals and alloys

Mesoscopic electronic transport through a disordered conductor can be described by a $N\times N$ transmission matrix $\hat{t}$ which relates the amplitudes of $N$ incoming and outgoing  transverse modes~\cite{1970_Landauer}. Dimensionless conductance is $g=\langle{\rm Tr}\left(\hat{t}^\dagger\hat{t}\right)\rangle=\sum_n\langle\tau_n\rangle$, where $\tau_n$ are the eigenvalues of the matrix $\hat{t}^\dagger\hat{t}$~\cite{1981_Fisher} and $\langle...\rangle$ denotes ensemble average. Therefore, electron transport in a metallic wire can be viewed as parallel transmission over $N$ orthogonal eigenchannels with individual transmissions of $\tau_n$. Due to the mesoscopic correlations~\cite{1994_Berkovits_Feng,1999_van_Rossum}, density of the transmission eigenvalues ${\cal D}(\tau)$ has a bimodal functional form~\cite{1984_Dorokhov,1986_Pichard_Eigenchannels,1987_Muttalib_Random_Matrix,1989_Mello,1994_Beenakker_exact,2005_Tian_Eigenvalues,2014_Gerardin_Wavefront_Shaping_Full_Matrix} with peaks at $\tau\rightarrow 0$ and $\tau\rightarrow 1$. The latter implies the existence of nearly perfect transmission eigenchannels~\cite{1984_Dorokhov,1986_Imry,1992_Pendry_math} that lead to e.g. universal conductance fluctuations~\cite{1985_Stone_UFC,1991_Altshuler} and quantum shot noise~\cite{1992_Beenakker_Shot_Noise,1994_Altshuler_Shot_Noise}. In Ref.~\cite{1994_Nazarov_Eigenvalues}, bimodal distribution was proven to be applicable to an arbitrary geometry of the conductor as long as the transport remains diffusive and free of dissipation.

The bimodal distribution obtained in the context of mesoscopic physics is also applicable to transport of classical waves in scattering media~\cite{2007_Akkermans_book}. In optics, rapid development of spatial light modulators has enabled an experimental access to the transmission eigenchannels~\cite{2012_Mosk_SLM_review} that led to experimental demonstration of enhanced transmission~\cite{2008_Vellekoop_Mosk,2012_Genack_Eigenvalues} with applications in focusing and imaging through turbid media~\cite{2008_Vellekoop_Mosk,2010_Popoff_Shaping_PRL,2010_Dogariu_Eigenchannels,2011_Katz_Silverberg_Wavefront_Shaping,2014_Popoff_Eigenvalues,2014_Cao_Open_Channels,2011_Stone_Coherent_Absorption,2013_Yu_Eigenchannels,2014_Cheng_Wavefront_Shaping_Energy_Deposition}. Absorption, common in optics, breaks energy conservation and makes the density of transmission eigenvalues~\cite{1998_Brouwer} as well as reflection~\cite{1996_Bruce_Chalker_Reflection_Eigenvalues,1996_Beenakker_random_laser,2013_Stone_Eigenvalues_with_Absorption} eigenvalues to depend on its strength. However, the questions of whether the geometry of the system could affect the eigenvalue density in dissipative systems and if so, how it would affect it, have not been addressed.

In this work we demonstrate that, unlike passive systems, the density of the transmission eigenvalues in absorbing disordered waveguides is geometry dependent, that is beyond predictions of the existing theory~\cite{1998_Brouwer}. This opens possibility of tuning the functional form of the eigenvalue density by choosing the shape of the boundary. Furthermore, we show that dissipation makes a profound impact on the densities of reflection eigenvalues $\rho$ and absorption eigenavlues $\alpha$, that can even depend on which side of the waveguide is being illuminated in the case of asymmetric waveguide shape. This is attributed to the fact that reflection matrices for illumination from different sides are no longer related in the presence of dissipation. Above a certain absorption threshold, the density of absorption eigenvalues exhibits a qualitative transformation from a single-peak to a double-peak function. The additional peak at $\alpha\simeq 1$ enables a nearly complete absorption at any frequency with an appropriate input wavefront. 

\begin{figure}[htbp]
\centering
\includegraphics[width=3.5in]{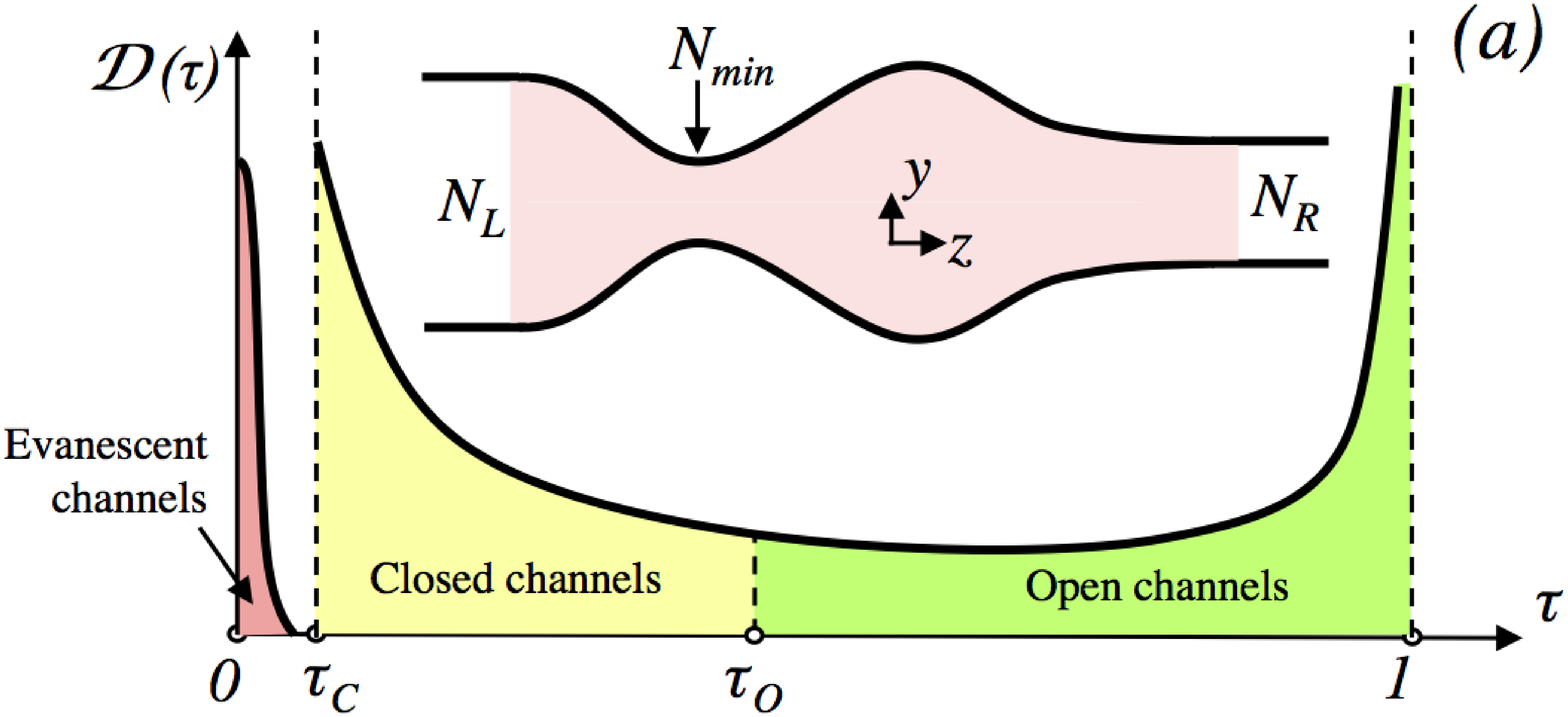}
\includegraphics[width=3.3in]{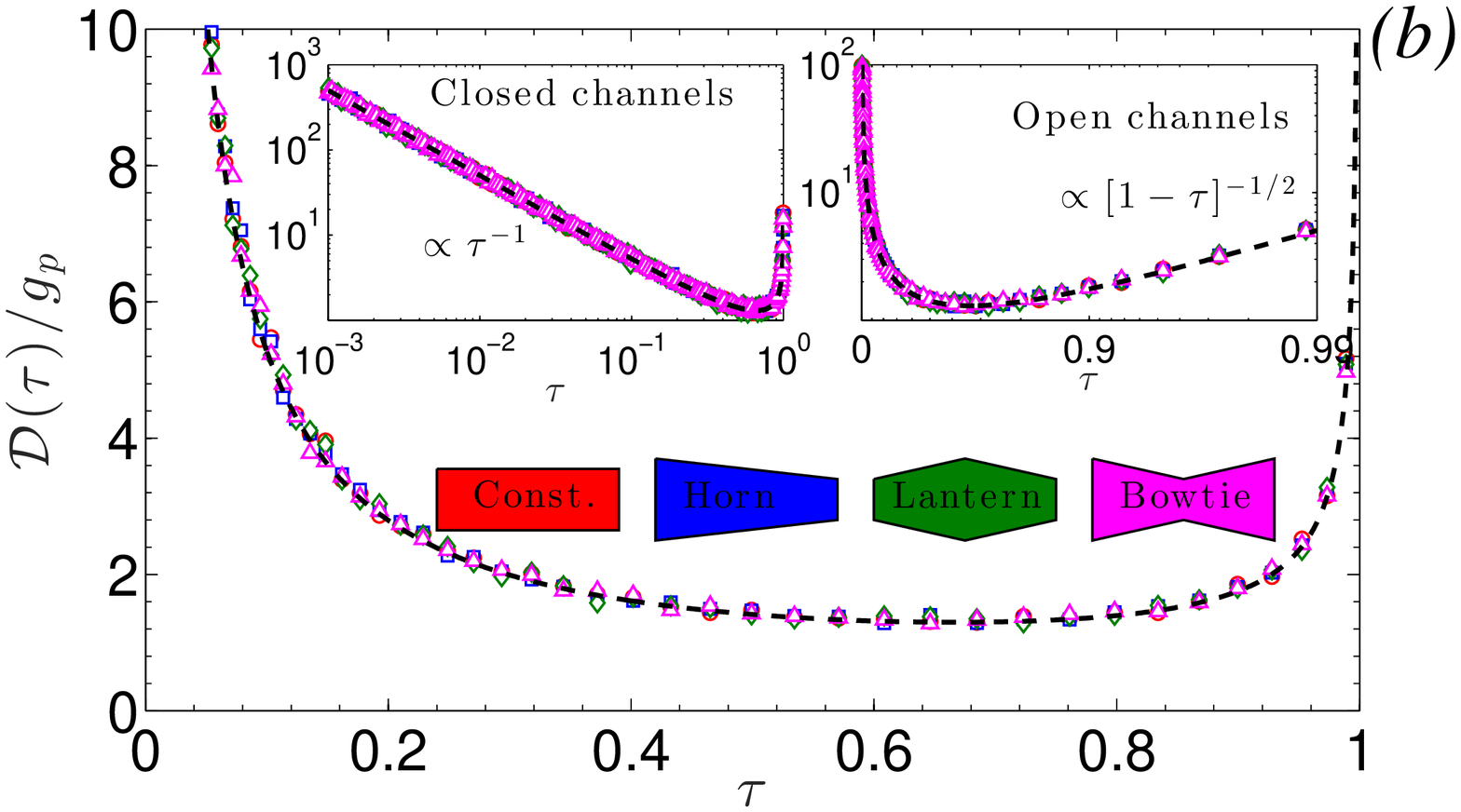}
\vskip -0.5cm
\caption{\label{fig:Poftau_passive} 
(Color online) (a) Schematic illustration of density of the transmission eigenvalues ${\cal D}(\tau)$ in a passive disordered waveguide of varying width $W(z)$ drawn in the inset. It is made up of open ($\tau_O<\tau\leq 1$), closed ($\tau_{C}<\tau<\tau_O$) and evanescent ($\tau<\tau_C$) eigenchannels.
(b) Normalized density of the transmission eigenvalues ${\cal D}(\tau)/g_p$ computed numerically for four passive waveguides shown. All data points fall onto the dashed line -- the bimodal distribution. Two insets show that the bimodal distribution correctly describes both $\tau\rightarrow 0$ (closed channels) and $\tau\rightarrow 1$ (open channels) limits, regardless of the waveguide shape.}
\end{figure}

{\it Transmission eigenvalues.}
We consider a variable width waveguide, schematically depicted in the inset of Fig.~\ref{fig:Poftau_passive}a, formed by reflecting boundaries at $y(z)=\pm W(z)/2$, where $W(z)$ is a smooth function of $z$. The leads on the left/right support $N_L$/$N_R$ propagating modes. The transport through the disordered region $0\leq z\leq L$ is described by a complex $N_R\times N_L$ matrix $\hat{t}$. For passive random media, density of the eigenvalues of matrix $\hat{t}^\dagger\hat{t}$ is ${\cal D}(\tau)=(g_p/2)\tau^{-1}(1-\tau)^{-1/2}$. In \cite{SI}, we reproduce this result using the circuit theory of Ref.~\cite{1994_Nazarov_Eigenvalues} with the dimensionless conductance given by $g_p^[W(z)]=(k\ell/2)[\int_{0}^{L}W^{-1}(z)dz]^{-1}$, where $k=2\pi/\lambda$ is the wavenumber, $\ell$ is the transport mean free path, and subscript {\it p} stands for ``passive''. For a waveguide with constant $W=N \, (\lambda/2)$ width we recover the well-known expression $g_p=(\pi/2)N\ell/L$~\cite{1997_Beenakker}. 

Figure ~\ref{fig:Poftau_passive}a schematically depicts ${\cal D}(\tau)$ with three contributions from open, closed and evanescent eigenchannels. Open ($\tau_O<\tau<1$) and closed ($\tau_C<\tau<\tau_O$) channels lead to the bimodal distribution. With $\tau_O\equiv[2e/(e^2+1)]^2\simeq 0.42$, the number of open eigenchannels is  $N_O=g_p$~\cite{1986_Imry,1992_Pendry_math}. The cutoff $\tau_C$ at the level of ballistic transmission~\cite{1984_Dorokhov,1997_Beenakker} is obtained by normalizing  $\int_{\tau_{C}}^{1}{\cal D}(\tau)d\tau$ to the number of propagating channels $N_{min}=W_{min}/(\lambda/2)$, see Fig.~\ref{fig:Poftau_passive}a. In a waveguide with a constriction, there are ${\rm min}(N_L,N_R)$ transmission eigenchannels, among which $N_E={\rm min}(N_L,N_R)-N_{min}$ are evanescent channels with intensity decaying on the scale of the wavelength inside the narrow portion of the waveguide and, therefore, $\tau\ll\tau_C$ for these channels~\cite{SI}. This boundary separating evanescent and closed channels is exaggerated for illustration in Fig.~\ref{fig:Poftau_passive}a, as in practice $\tau_{C}\simeq 0$. 

%%%%%%%%%%%%%%%%%%%%%%%%%%%%%%%%%%%%%%%%%%%%%%%%%%%%%%%%%%%%%%%%
%%%%%%%%%%%%%%%%%%%%%%%%%%%%%%%%%%%%%%%%%%%%%%%%%%%%%%%%%%%%%%%%

\begin{figure}[htbp]
\centering
\includegraphics[width=3.5in]{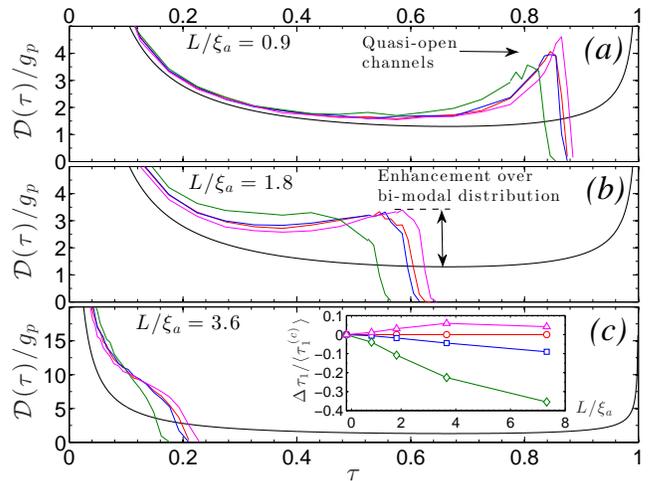}
\vskip -0.3cm
\caption{\label{fig:Poftau} 
(Color online) Density of the transmission eigenvalues ${\cal D}(\tau)/g_p$ in absorbing diffusive waveguides depends not only on the absorption strength but also on the confinement geometry. Four colored curves correspond to four waveguides with matched color in Fig.~\ref{fig:Poftau_passive}b. Absorption strength is $L/\xi_a=0.9$ -- (a), $1.8$ -- (b), and $3.6$ -- (c). The universal bimodal distribution of passive waveguides (solid line) is shown for reference. Inset of (c): normalized deviation of maximum transmission eigenvalue $\Delta\tau_1=\langle\tau_{1}\rangle-\langle\tau_{1}^{(c)}\rangle$ in four waveguides of different shape from that in the constant width waveguide $\langle\tau_{1}^{(c)}\rangle$, as a function of absorption $L/\xi_a$.}
\end{figure}

Applicability of the bimodal distribution for open and closed channels is confirmed in Fig.~\ref{fig:Poftau_passive}b. It shows ${\cal D}(\tau)/g_p$ computed numerically using Kwant simulation package~\cite{2014_Groth_Kwant} (see \cite{SI} for details) for four waveguides of different shape (drawn in the inset): constant width waveguide of width $W=273\times(\lambda/2)$; horn waveguide of width linearly decreasing from $W_L=400\times(\lambda/2)$ to $W_R=200\times(\lambda/2)$; lantern waveguide of width linearly tapered from $W_M=400\times(\lambda/2)$ in the middle to $W_L=W_R=200\times(\lambda/2)$ at the two ends; and bowtie of width tapered from $W_L=W_R=400\times(\lambda/2)$ at the ends to $W_M=200\times(\lambda/2)$ in the middle. The conductance in the four systems is $g_p=13.9,\ 14.2,\ 13.5$ and $13.9$ respectively. The other system parameters are $L/\ell\simeq 31$, $k\ell\simeq 60$. Fig.~\ref{fig:Poftau_passive}b clearly shows that the bimodal distribution, including the asymptotes for $\tau\rightarrow 0,1$ in insets of Fig.~\ref{fig:Poftau}, describes open and closed eigenchannels in waveguides of different shapes without any fitting parameters. The nonuniversal contribution of evanescent channels  to ${\cal D}(\tau\simeq 0)$ cannot be clearly distinguished from the peak of closed channels in the numerical data. Nevertheless, the evanescent channels can make up a substantial fraction of the total channels, e.g., in the bowtie waveguide, one half of the transmission eigenchannels are evanescent and have the vanishingly small values of $\tau$.

%%%%%%%%%%%%%%%%%%%%%%%%%%%%%%%%%%%%%%%%%%%%%%%%%%%%%%%%%%%%%%%%
%%%%%%%%%%%%%%%%%%%%%%%%%%%%%%%%%%%%%%%%%%%%%%%%%%%%%%%%%%%%%%%%

Absorption breaks flux conservation and time-reversal symmetry, leaving optical reciprocity the only constraint on the scattering matrix $\hat{S}$ of the system~\cite{2000_Carminati_Reciprocity}. In \cite{SI} we show that it relates (in each realization of disorder) the transmission matrices for waves incident from the left and right as $\hat{t}^T=\hat{t}^\prime$, where superscript $T$ denotes matrix transpose. This relationship signifies that even in the presence of absorption, $\hat{t}^\dagger\hat{t}$ and $\hat{t}^{\prime\dagger}\hat{t}^\prime$ have the same set of non-zero eigenvalues. In waveguides of constant width, density of the transmission eigenvalues has been predicted to depend on the strength of absorption and an analytical expression for ${\cal D}(\tau)$ has been derived in the limit of strong absorption $L/\xi_a\gg 1$~\cite{1998_Brouwer}. However, the possibility that ${\cal D}(\tau)$ can depend on the waveguide shape has not been explored.

Figures~\ref{fig:Poftau}a-c show density of the transmission eigenvalues for waveguides of different shape with three values of absorption: $L/\xi_a=0.9,\ 1.8,$ and $3.6$. $\xi_a=[\ell\ell_a/2]^{1/2}$ is the diffusive absorption length and $\ell_a$ is the ballistic absorption length. Common to all geometries, $\tau\simeq 1$ eigenvalues are attenuated so that the density no longer reaches unity. Instead, the maximum eigenvalue $\langle\tau_{1}\rangle<1$. Open channels are redistributed throughout $\tau_C<\tau<{\rm max}(\tau_{1})$ interval so that the eigenvalue density is consistently higher than that in passive systems. However, unlike the bimodal distribution for the passive systems, see Fig.~\ref{fig:Poftau_passive}b, ${\cal D}(\tau)$ {\it is no longer universal and exhibits a clear shape dependence}. The maximum transmission eigenvalue is lowest for the lantern geometry. Such behavior can be understood as the narrower openings and slanted walls of the lantern waveguide reduce the escape probability and increase the effective absorption, leading to smaller $\langle\tau_{1}\rangle$. In contrast, the situation is reversed in the bowtie waveguide, see Fig.~\ref{fig:Poftau}. This structure has wider openings and, therefore, waves are more likely to escape without being strongly attenuated. The normalized deviation of the largest eigenvalue $\langle\tau_{1}\rangle$ in waveguides of different shapes from that in the constant width waveguide, $\langle\tau_{1}^{(c)}\rangle$, is plotted in the inset of Fig.~\ref{fig:Poftau}c. The deviation increases with absorption strength and can be either negative (horn, lantern) or positive (bowtie). However, at the largest value of absorption of $L/\xi_a\simeq 7.3$, the deviation is reduced in the bowtie waveguide, which can be understood as follows. For strong absorption $L\gg\xi_a$, short propagation paths dominate transport~\cite{2014_Cao_Open_Channels}, so we expect the deviation to decrease in this limit because all geometries have the same length $L$. Such ballistic-like propagation is more favored due to the constriction in the bowtie waveguide, where this transition occurs first.

%%%%%%%%%%%%%%%%%%%%%%%%%%%%%%%%%%%%%%%%%%%%%%%%%%%%%%%%%%%%%%%%
\begin{figure}[htbp]
\centering
\vskip -0.2cm
\includegraphics[width=3.3in]{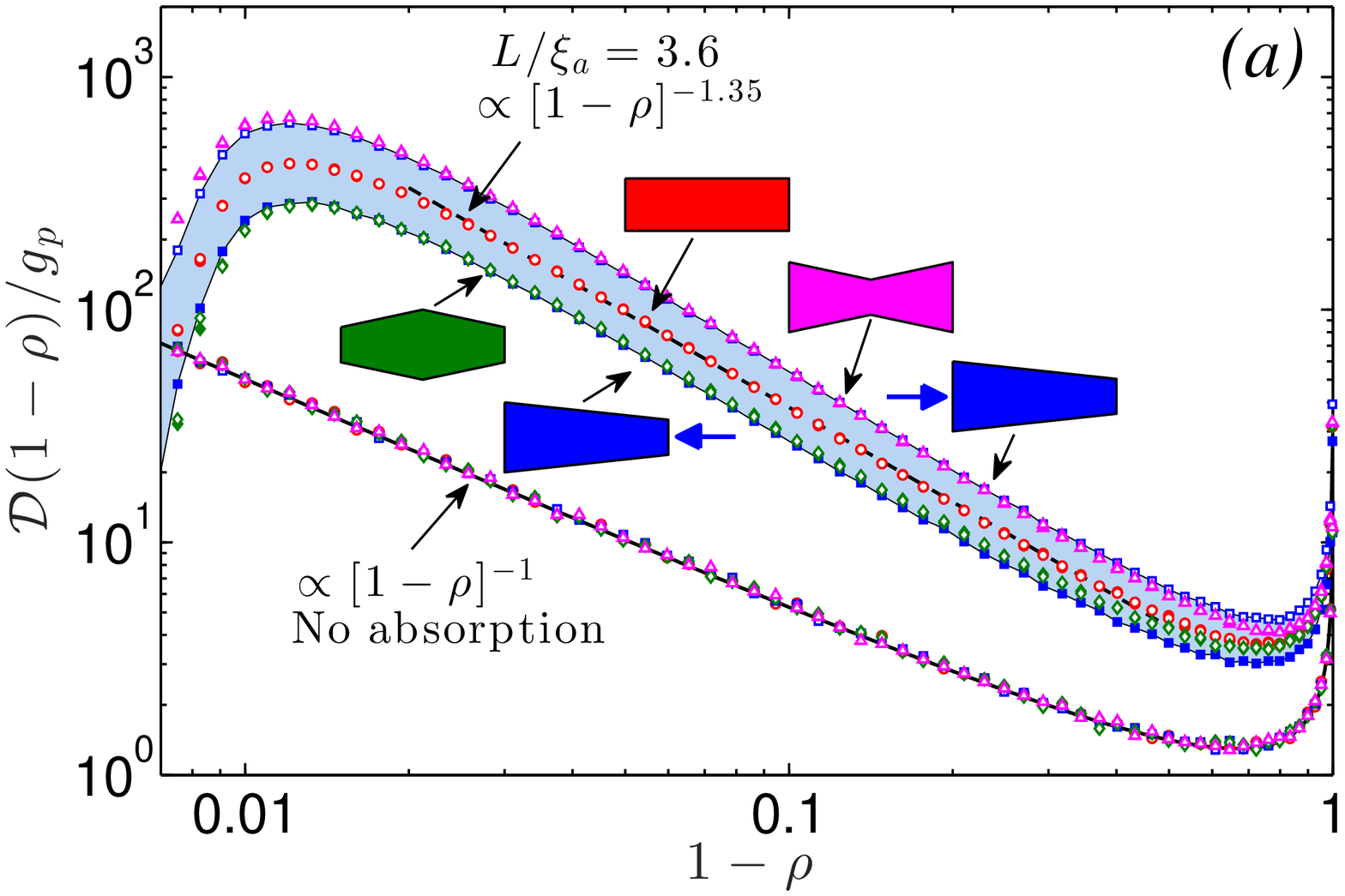}
\vskip -0.2cm
\includegraphics[width=3.3in]{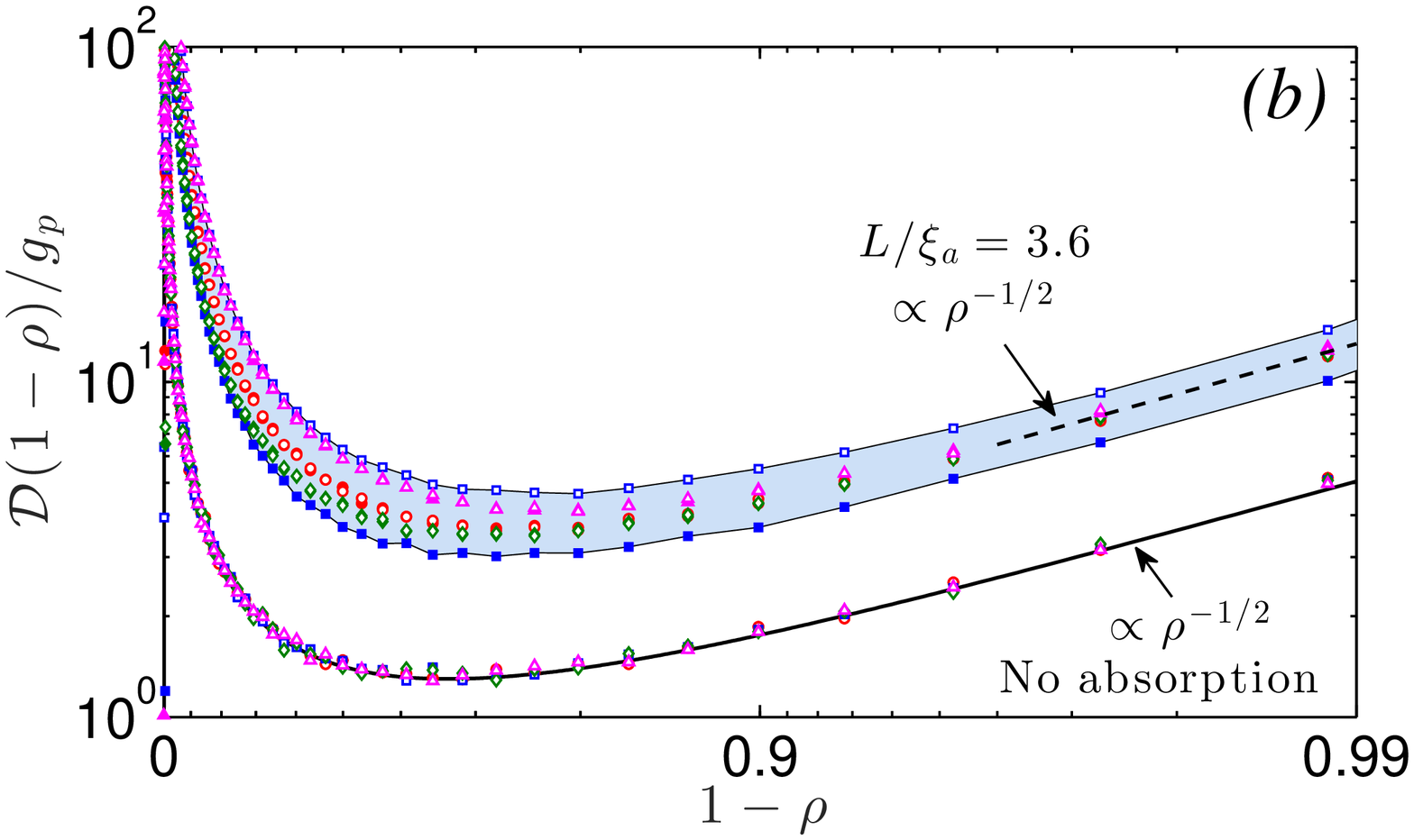}
\vskip -0.3cm
\caption{\label{fig:Pofrho} 
(Color online) Density ${\cal D}(1-\rho)/g_p$ of the reflection eigenvalues $\rho$ in diffusive waveguides of different shape with absorption ($L/\xi_a=3.6$) and without absorption. Panels (a,b) show the power scaling behaviors at $(1-\rho)\rightarrow 0$ and $1$ respectively. Without absorption all eigenvalue densities, regardless of the waveguide shape or input direction, fall onto the bimodal distribution -- solid curve. In absorbing systems, ${\cal D}(1-\rho)/g_p$ obtained for waves incident from the left/right are shown with open/filled symbols. For all symmetric waveguides (constant width, lantern, bowtie) with absorption, $N_L=N_R$, filled and open symbols coincide. For the asymmetric horn waveguide ($N_L\neq N_R$), a large disparity between left/right illumination is highlighted by shaded area.}
\end{figure}
%%%%%%%%%%%%%%%%%%%%%%%%%%%%%%%%%%%%%%%%%%%%%%%%%%%%%%%%%%%%%%%%

{\it Reflection eigenvalues.}
In a passive system, the energy conservation and symmetry requirements make all non-zero eigenvalues of $\hat{t}^\dagger\hat{t}$, $\hat{I}-\hat{r}^\dagger\hat{r}$, $\hat{t}^{\prime\dagger}\hat{t}^\prime$, $\hat{I}-\hat{r}^{\prime\dagger}\hat{r}^\prime$ identical, where $\hat{r}$ ($\hat{r}^\prime$) represents the reflection matrix for waves incident from left (right) end of the waveguide~\cite{SI}. This leads to the bimodal distribution of the density of $1-\rho$ for both left and right reflection eigenvalues $\rho$ and regardless of the shape of the waveguide. In an asymmetric waveguide with $N_L\neq N_R$ (we will assume $N_L>N_R$ without loss of generality), the $N_L\times N_L$ matrix $\hat{r}^\dagger\hat{r}$ also has $N_L-N_R$ eigenvalues with $\rho =  1$, giving the {\it perfectly reflecting eigenchannels} for light incident from the left (wider opening). Meanwhile, for waves incident from the right (narrower opening), there are no perfectly reflecting eigenchannels because $N_R\times N_R$ matrix $\hat{r}^{\prime\dagger}\hat{r}^\prime$ has only $N_R$ eigenvalues, all of which have corresponding transmission eigenvalues that are non-zero. The results of the numerical simulations in passive waveguides of different shape, c.f. Fig.~\ref{fig:Pofrho}, confirm that the density of both left/right reflection eigenvalues ${\cal D}(1-\rho)$ follows the universal bimodal distribution, which still holds in asymmetric waveguides as the perfectly reflecting eigenchannels only have a singular contribution at $\rho = 1$.

Due to absence of flux conservation in systems with absorption, the links between reflection and transmission matrices {\it and} between left/right reflection matrices are severed~\cite{SI}. Consequently, in each disorder realization, the eigenvalues of $\hat{r}^\dagger\hat{r}$ and $\hat{r}^{\prime\dagger}\hat{r}^\prime$ are not necessarily identical and they are no longer related to the transmission eigenvalues. Our numerical simulations confirm that the perfect reflecting channels are removed by absorption as {\it all} reflection eigenvalues become less than unity. Furthermore, in asymmetric waveguides ($N_L\neq N_R$), the densities of reflection eigenvalues differ for waves incident from left/right side of the waveguide, as shown in Figs.~\ref{fig:Pofrho}a,b for the horn geometry. Even for symmetric waveguides ($N_L=N_R$), ${\cal D}(\rho)$ is still clearly shape-dependent, as seen in Figs.~\ref{fig:Pofrho}a,b for the constant width, lantern and bow-tie geometries: ${\cal D}(1-\rho)$ are distinctly different in $(1-\rho)\rightarrow 0$ limit while in the limit $(1-\rho)\rightarrow 1$ the difference is greatly reduced.

Figs.~\ref{fig:Pofrho}b shows that power exponent in ${\cal D}(1-\rho)\propto \rho^{-1}$ for $(1-\rho)\rightarrow 1$ is independent of the waveguide shape/input direction and it is the same as in a passive system. For $(1-\rho)\rightarrow (1-\rho_{max})$, we find that the power exponent in ${\cal D}(1-\rho)\propto (1-\rho)^{-1.35}$ has a weak shape dependence. The value $1.35$ is smaller than $3/2$ found in Refs.~\cite{1996_Bruce_Chalker_Reflection_Eigenvalues,1996_Beenakker_random_laser} for $a=N\ell/\ell_a\gg 1$ in constant width waveguides. We attribute the discrepancy to insufficiently large value of $a=1.9$ for the case shown in Fig.~\ref{fig:Pofrho}a. Finally, we note that although the effect of the absorption on the density of reflection eigenvalues is known for the constant width waveguides~\cite{1996_Bruce_Chalker_Reflection_Eigenvalues,1996_Beenakker_random_laser}, the existing theory cannot explain shape/input direction dependence of ${\cal D}(\rho)$. This calls for further theoretical investigations.

%%%%%%%%%%%%%%%%%%%%%%%%%%%%%%%%%%%%%%%%%%%%%%%%%%%%%%%%%%%%%%%%
\begin{figure}[htbp]
\centering
\vskip -0.2cm
\includegraphics[width=3.3in]{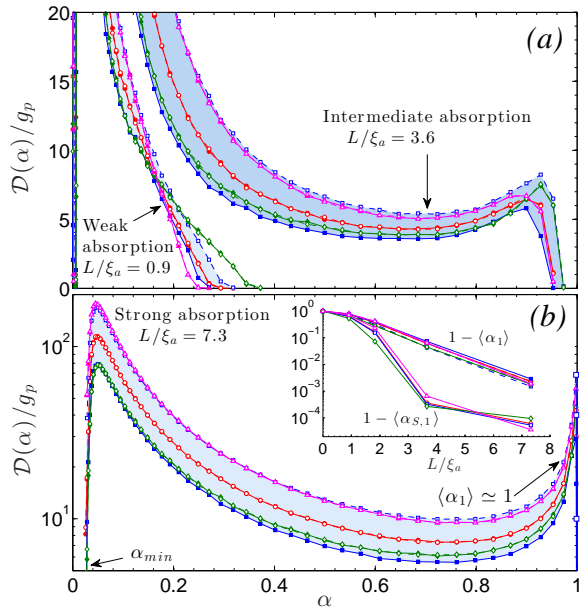}
\vskip -0.3cm
\caption{\label{fig:Pofabs} 
(Color online) Density ${\cal D}(\alpha)$ of absorption eigenvalues $\alpha$ in disordered waveguides of different shape, under on-sided illumination, evolves from one peak function at weak absorption (the ensemble-averaged maximum absorption eigenvalue $\langle\alpha_{1}\rangle\ll 1$) to double peak function at intermediate absorption ( $\langle\alpha_{1}\rangle\lesssim 1$)  in panel a, and the second peak moves to $\alpha \simeq 1$ at strong absorption ($\langle\alpha_{1}\rangle\simeq 1$)  in panel b. Symbol notations are the same as in Fig.~\ref{fig:Pofrho}. In all cases the normalized density of absorption eigenvalues exhibits strong dependence on the shape of the waveguide, and for asymmetric (horn) waveguide also on the input direction. The inset in panel (b) plots the ensemble-averaged maximum absorption eigenvalue $\langle\alpha_{1}\rangle$ vs. the absorption strength $L/\xi_a$. For comparison, the maximum absorption eigenvalue $\langle\alpha_{S,1}\rangle$ for two-sided illumination are also shown.}
\end{figure}
%%%%%%%%%%%%%%%%%%%%%%%%%%%%%%%%%%%%%%%%%%%%%%%%%%%%%%%%%%%%%%%%

{\it Absorption eigenvalues.}
In a dissipative system, the non-unitary part of the scattering matrix $\hat{I}-\hat{S}^\dagger\hat{S} \equiv \hat{A}_S$ accounts for absorption~\cite{1998_Beenakker_Mesossopic_Statistics_of_RL} and its largest eigenvalue $\alpha_{S,1}$ tells the maximum absorption that can be achieved by shaping the input wavefront~\cite{2011_Stone_Coherent_Absorption}. This requires controlling all modes incident onto both sides of the waveguide. However, more common in experiments is only one side of the system illuminated. In such case the matrix $\hat{A}=\hat{I}-\hat{r}^\dagger\hat{r}-\hat{t}^\dagger\hat{t}$ describes the absorption of input light. Its largest eigenvalue $\alpha_1$ determines the maximum absorption in a given system when only one side is accessible. Similar to density of the reflection eigenvalues, ${\cal D}(\alpha)$ depends on the shape of the waveguide and the input direction, c.f. Fig.~\ref{fig:Pofabs}a,b. Common to all geometries, the functional form of ${\cal D}(\alpha)$ undergoes a qualitative change with an increase of absorption strength. At weak absorption, the eigenvalue density monotonously decreases toward zero with an increase of $\alpha$, c.f. Fig.~\ref{fig:Pofabs}a. At the increased absorption, the density  develops the second maximum at $\alpha\simeq 1$. Even in this regime, there exists an upper bound, which approaches unity exponentially, c.f. inset of \ref{fig:Pofabs}b. A coherent perfect absorber proposed in Ref.~\cite{2010_Stone_Coherent_Absorber} achieves $100\%$ absorption but requires full control of incident wavefront and a specific amount of absorption. In contrast, we show that at any frequency and with any absorption (above a certain threshold) the maximum achievable absorption with one-sided excitation $\alpha_1$ can be close to unity. Moreover, with the left end of the waveguide being illuminated, for example, we can achieve nearly perfect absorption by controlling a fraction $N_L/(N_L+N_R)$ of all input channels, that can be small in e.g. a horn waveguide with $N_L<N_R$. 

We note that absorption dependence of the maximum eigenvalue $\langle\alpha_1\rangle$ for one-sided illumination is qualitatively different from $\langle\alpha_{S,1}\rangle$ for two-sided illumination, c.f. inset in Fig.~\ref{fig:Pofabs}b. The former approaches unity exponentially, $1 - \langle \alpha_1 \rangle \propto \exp[-L/\xi_a]$. In contrast, excitation from both sides results in a sharp transition at $L/\xi_a\sim 3$, above which strong enhancement of absorption~\cite{2011_Stone_Coherent_Absorption} with $\langle\alpha_{S,1}\rangle\simeq 1$ becomes possible. The critical value of absorption can be estimated by comparing the diffusion time without absorption $L^2/D\pi^2$ to the absorption time $t_a=\xi_a^2/D$, where $D$ is the diffusion coefficient. Equating these two characteristic time scales results in $L/\xi_a=\pi$ which agrees with Fig.~\ref{fig:Pofabs}b. This offers an absorption analogy with diffusive random laser~\cite{1994_Lawandy_nature,2003_Cao_Progress_in_Optics,2005_Yamilov_correlations} where exactly the same amount of gain corresponds to the lasing threshold, giving {\it output} to all sides.

{\it Conclusions.} 
We present a numerical study on the densities of the transmission, reflection and absorption eigenvalues in diffusive absorbing waveguides of various shapes. Without absorption, the density of transmission eigenvalues is given by the universal bimodal distribution, independent of both microscopic and macroscopic details of the system. We show that in the presence of absorption the densities of transmission as well as reflection and absorption eigenvalues depend on the confinement geometry of random media. In addition, absence of flux conservation in lossy media breaks the relationship between reflection (and absorption) matrices with incident waves from the opposite sides. As a consequence, we demonstrate that the density of reflection and absorption eigenvalues can depend on the incident direction. Therefore, the geometry of the system represents an additional degree of freedom in coherent control of wave transmission, reflection and absorption in turbid media. This offers design opportunities not available in passive random media. The existing theoretical description of the eigenchannels in dissipative random media cannot account for the effects from the geometry of the system. We believe that this work will stimulate a new body of research, with profound implications for waves of different nature where dissipation is common, e.g., the microwave, acoustic wave, etc.

{\it Acknowledgments. }
We thank A.~D.~Stone, P.~Brouwer, C.~W.~J.~Beenakker for stimulating discussions and B.~van~Heck for technical help with Kwant simulation package. This work is supported by the National Science Foundation under Grants No. DMR-1205307 and DMR-1205223. 

%\bibliography{../../Bibliography/latex_bibliography}
%\bibliography{../../../Bibliography/latex_bibliography}
\bibliography{2015_Eigenchannels0.bbl}

\end{document}

\title{Supplementary Information: \\
Shape-dependence of transmission, reflection and absorption eigenvalue densities \\
in disordered waveguides with dissipation}

\author{A. Yamilov}
\email{yamilov@mst.edu}
\affiliation{Department of Physics, Missouri University of Science and Technology, Rolla, Missouri 65409,USA}
\author{S. Petrenko}
\affiliation{Department of Physics, Missouri University of Science and Technology, Rolla, Missouri 65409,USA}
\author{R. Sarma}
\affiliation{Department of Applied Physics, Yale University, New Haven, CT, 06520, USA}
\author{H. Cao}
\email{hui.cao@yale.edu}
\affiliation{Department of Applied Physics, Yale University, New Haven, CT, 06520, USA}
\maketitle

%%%%%%%%%%%%%%%%%%%%%%%%%%%%%%%%%%%%%%%%%%%%%%%%%%%%%%%%%%%%%%%%

We consider a two-dimensional (2D) waveguide formed by reflecting boundaries at $y(z)=\pm W(z)/2$. The width of the waveguide $W(z)$ is a smooth function of $z$ - the axial coordinate, see Fig.~\ref{fig:circuit_theory}. The waveguide is coupled to two leads (empty waveguides) at $z = 0$ and $z = L$.  The leads have constant width, $W_L = W(0)$ and $W_R = W(L)$, thus supporting $N_L=W_L/(\lambda/2)$ and $N_R=W_R/(\lambda/2)$ guided modes respectively. The transport through the disordered region $0\leq z\leq L$ is assumed to be diffusive, i.e., $\ell\ll L\ll \xi$, where $\ell$ and $\xi$ are transport mean free path and the localization length respectively, and the dimensionless conduction $g_p\gg 1$, where subscript {\it p} stands for ``passive''. 

\subsection{Derivation of density \\ of the transmission eigenvalues: Circuit theory \label{subsec:circuit_theory}}

%%%%%%%%%%%%%%%%%%%%%%%%%%%%%%%%%%%%%%%%%%%%%%%%%%%%%%%%%%%%%%%%
\begin{figure}[htbp]
\centering
\vskip -0.3cm
\includegraphics[width=2.5in]{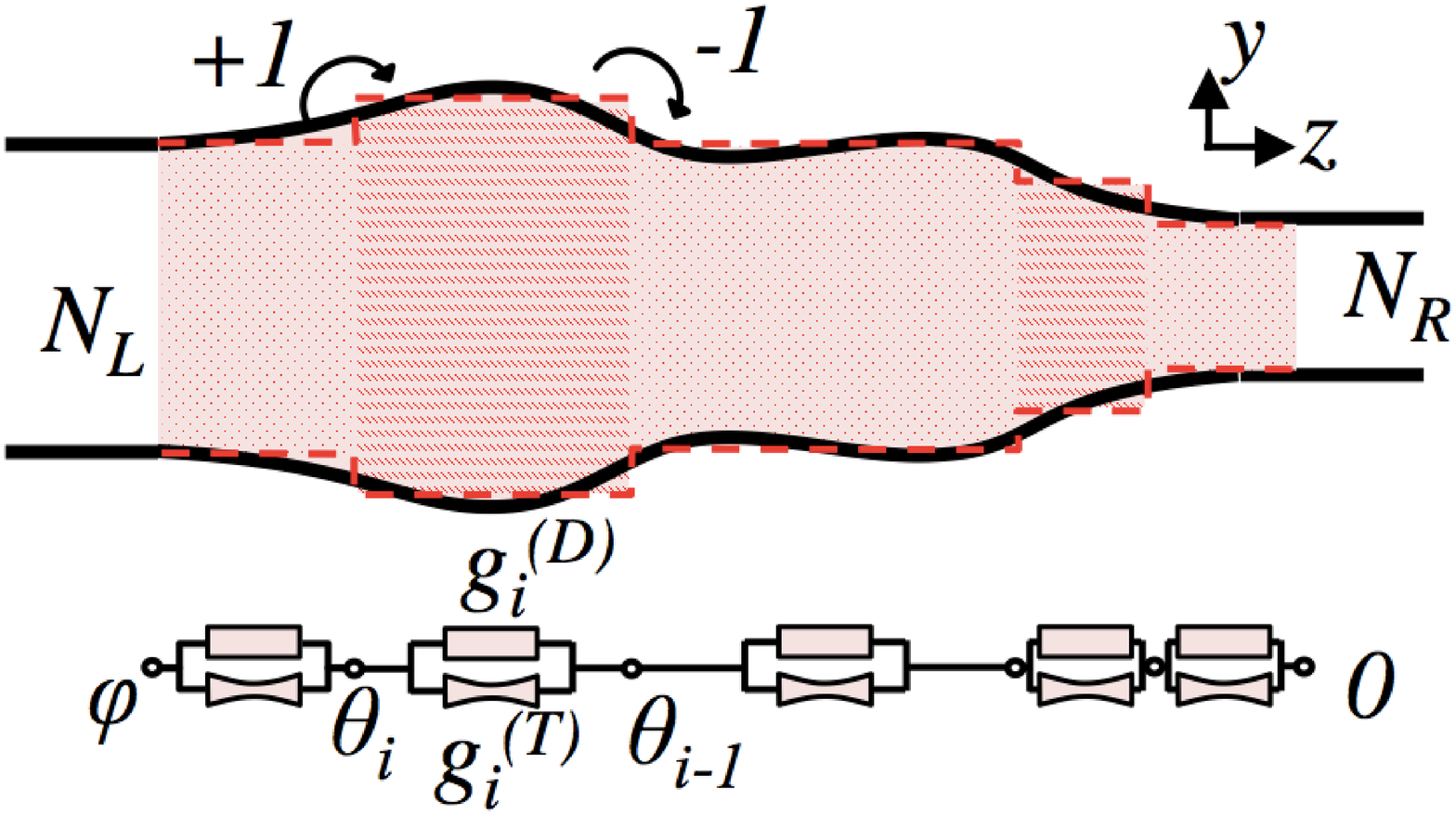}
\vskip -0.2cm
\caption{\label{fig:circuit_theory} Schematic representation of a disordered waveguide with varying width $W(z)$. It is approximated by a sequence of segments, each having a  constant width $W_i$, with $W_i-W_{i-1}=\pm\lambda/2$ corresponding to adding/removing one waveguide mode with real propagation constant. Circuit theory~\cite{1999_Nazarov_Eigenvalues} representation of the waveguide in terms of diffusive conductors (propagating modes) and tunneling junctions (evanescent modes) is depicted below.}
\end{figure}
%%%%%%%%%%%%%%%%%%%%%%%%%%%%%%%%%%%%%%%%%%%%%%%%%%%%%%%%%%%%%%%%

In this section, we derive the bimodal distribution for the density of transmission eigenvalues in passive disordered waveguides of an arbitrary shape. Circuit theory was developed in Ref.~\cite{1999_Nazarov_Eigenvalues} in context of mesoscopic transport of electrons in disordered conductors and tunneling junctions. However, the methodology is general and applicable to any system that exhibits phase coherent wave transport. Below we will apply it to light transport in a disordered waveguide to derive the density of transmission eigenvalues. The first step is to represent a complex system as a network of basic elements, in our case, diffusive conductors (resistors) and tunneling junctions, as follows. 

As depicted in Fig.~\ref{fig:circuit_theory}, a waveguide of variable width can be represented as a sequence of segments $z_{i-1}$$<z<$$z_{i}$ in which the number of waveguide modes (with real propagation constant) is $N_i={\rm floor}[W(z)/(\lambda/2)]$. The steps of $\Delta W=W_i-W_{i-1}=\lambda/2$ determine the length of each segment, and the number of waveguide modes in consecutive segments differs by $\pm1$. $1$ ($-1$) corresponds to the conversion of an evanescent (propagating) mode with imaginary (real) valued propagation constant to the propagating (evanescent) one. 

For a unified description of all segments, we assume each segment has $N_{max}=\max[W(z)]/(\lambda/2)$ modes. Hence the $i$'th segment has $N_i$ propagating and $N^{(e)}_i$ evanescent modes so that $N_i+N^{(e)}_i=N_{max}$ for any $i$. To complete mapping onto a circuit network, see Fig.~\ref{fig:circuit_theory}, we model wave transport via $N_i$ propagating modes in the $i$'th segment as a diffusive conductor with dimensionless conductance $g^{(D)}_i\gg 1$ and $N^{(e)}_i$ evanescent modes as a tunneling junction with very small conductance $g^{(T)}_i\ll 1$. In fact, the evanescent modes~\cite{1990_Bagwell} are not usually considered in theoretical models~\cite{1997_Beenakker} because their contributions to the overall transport are negligible~\cite{2007_Froufe-Perez_PRE,2013_Yamilov_Closed_Channels}. Lastly we note that the diffusive conductor and tunneling junction are connected in parallel for each segment whereas the successive segments are connected in series. 

Next we introduce the ``phase'' and the ``matrix current'' in such a network. 
The matrix current ${\cal I}$ is analogous to the current, and the phase $\phi$ to the electric potential in an electronic circuit, thus the phase drop across a circuit element determines the matrix current through it. Current-phase relationships are given by ${\cal I}^{(D)}_i=g^{(D)}_i(\theta_i-\theta_{i-1})$ and ${\cal I}^{(T)}_i=g^{(T)}_i\sin(\theta_i-\theta_{i-1})$ for the diffusive conductor and tunneling junction respectively~\cite{2009_Nazarov_Book_Quantum_Transport}.
${\cal I}$ obeys Kirchhoff's rules which for our system in Fig.~\ref{fig:circuit_theory} gives ${\cal I}(\phi)={\cal I}_i(\theta_i-\theta_{i-1})={\cal I}^{(D)}_i(\theta_i-\theta_{i-1})+{\cal I}^{(T)}_i(\theta_i-\theta_{i-1})$. ${\cal I}^{(D)}_i$ and ${\cal I}^{(T)}_i$ represents matrix current through $i$'th diffusive conductor and tunneling junction respectively, and $\theta_i$ is the phase at node $i$. 
Applying these rules we find
\begin{equation}
{\cal I}(\phi)=g^{(CT)}\phi \, , 
\label{eq:matrix_current1}
\end{equation}
with
\begin{equation}
\frac{1}{g^{(CT)}}=\sum\limits_{i=1}^{M} \frac{1}{g^{(D)}_i+g^{(T)}_i} \simeq \sum\limits_{i=1}^{M} \frac{1}{g^{(D)}_i} \, ,
\label{eq:matrix_current2}
\end{equation}
where $M$ is the number of segments, and the diffusive conductance of the $i$-th segment $g_i^{(D)}=(\pi/2)N_i\ell/(z_{i}-z_{i-1})$ is much larger than $g^{(T)}_i$. 

The density of transmission eigenvalues follows from the relationship between ${\cal I}(\phi)$ and $\phi$ drop across the entire system~\cite{2009_Nazarov_Book_Quantum_Transport}:
\begin{equation}
{\cal D}(\tau)=\frac{1}{2\pi\tau\sqrt{1-\tau}}{\rm Re}\left[{\cal I}\left(\pi-0+2i{\rm arccosh}(\tau^{-1/2})\right)\right].
\label{eq:Poftau_circuit_theory_definition}
\end{equation}
Substituting Eqs.~(\ref{eq:matrix_current1},\ref{eq:matrix_current2}) into Eq.~(\ref{eq:Poftau_circuit_theory_definition}) we obtain the bimodal distribution with dimensionless conductance $g_p$ given by Eq.~(\ref{eq:matrix_current2}). In the continuous limit, the summation over the segments in Eq.~(\ref{eq:matrix_current2}) is replaced with an integral as
\begin{equation}
g_p^{(CT)}=(k\ell/2)\left[\int_{0}^{L}W^{-1}(z)dz\right]^{-1} .
\label{eq:gcp}
\end{equation}

\subsection{Scattering matrix of dissipative waveguides with varying cross-section}

Scattering matrix $\hat{S}$ contains full information about the transport properties of a system. It is related to the transmission, reflection matrices for light incident from the left $\hat{t},\ \hat{r}$ and right $\hat{t}^\prime,\ \hat{r}^\prime$ as follows:
\begin{equation}
\vec{a}^{\ out}=
\left( 
\begin{array}{c}
\vec{a}^{\ out}_{L}\\
\vec{a}^{\ out}_{R}\end{array} 
\right)
=\hat{S}\vec{a}^{\ in}=
\left( 
\begin{array}{cc}
\hat{r} & \hat{t}^\prime \\
\hat{t} & \hat{r}^\prime \end{array} 
\right)
\left( 
\begin{array}{c}
\vec{a}^{\ in}_{L}\\
\vec{a}^{\ in}_{R}\end{array} 
\right).
\label{eq:S}
\end{equation}
$\vec{a}^{\ in}_{L}$ and $\vec{a}^{\ out}_{L}$ are $N_L\times 1$ vectors for amplitudes of fields in the incoming/outgoing modes at $z<0$, and $\vec{a}^{\ in}_{R}$ and $\vec{a}^{\ out}_{R}$ are $N_R\times 1$ vectors for amplitudes of the incoming/outgoing modes at $z>L$. The matrices are, in general, rectangular: $\hat{t}$ -- $N_R\times N_L$, $\hat{r}$ -- $N_L\times N_L$, $\hat{t}^\prime$ -- $N_L\times N_R$, $\hat{r}^\prime$ -- $N_R\times N_R$, and $\hat{S}$ -- $(N_L+N_R)\times(N_L+N_R)$. In passive systems, due to conservation of flux ($\hat{S}^\dagger\hat{S}=\hat{I}$), time-reversal symmetry ($\hat{S}^*\hat{S}=\hat{I}$), and optical reciprocity ($\hat{S}^{T}=\hat{S}$), the non-zero eigenvalues are identical for $\hat{t}^\dagger\hat{t}$, $\hat{I}-\hat{r}^\dagger\hat{r}$, $\hat{t}^{\prime\dagger}\hat{t}^\prime$, $\hat{I}-\hat{r}^{\prime\dagger}\hat{r}^\prime$ matrices, i.e., $\tau = 1 - \rho = \tau^{\prime} = 1 - \rho^{\prime}$. The number of non-zero eigenvalues in a disordered waveguide is given by ${\rm min} (N_L, N_R)$.

A distinct group of evanescent eigenchannels arises in waveguides with a constriction $N_L,N_R>N_{min}$. This can be traced to the fact that the number of waveguide modes with real propagation constants is width dependent: $N(z)=W(z)/(\lambda/2)$. A constriction reduces the number of propagating modes and causes the rest to become evanescent. Although these modes can still transfer energy via tunneling and scattering in and out of the propagating modes, the inefficient transport~\cite{2007_Froufe-Perez_PRE,2013_Yamilov_Closed_Channels} results in exponential attenuation of their amplitudes $\sim\exp[-d/\lambda]$. The consequence of the attenuation can be better understood in terms of transfer matrix $\hat{M}_i$ of each segment of the waveguide, see Fig.~\ref{fig:circuit_theory}. Unlike  $\hat{S}$ matrix, the transfer matrix relates the incoming and outgoing waves on the left to those on the right  (compare to Eq.~\ref{eq:S})
\begin{equation}
\left( 
\begin{array}{c}
\vec{a}^{\ out}_{R}\\
\vec{a}^{\ in}_{R}\end{array} 
\right)
=\hat{M}
\left( 
\begin{array}{c}
\vec{a}^{\ in}_{L}\\
\vec{a}^{\ out}_{L}\end{array} 
\right).
\label{eq:M}
\end{equation}
which makes it multiplicative $\hat{M}=\Pi_i \hat{M}_{i=1}^{M}$. It is now easy to see that a set of small eigenvalues in one segment leads to small eigenvalues in the overall matrix $\hat{M}$. This observation can be restated mathematically as following: the rank of a product cannot exceed the smallest rank of the multiplier matrices~\cite{1999_Magnus_Matrix_Product}. Because the transfer and transmission matrices are related ~\cite{1990_Pichard_Eigenchannels,1997_Beenakker} via
\begin{equation}
\left[ 1+\hat{M}\hat{M}^\dagger+\left(\hat{M}\hat{M}^\dagger\right)^{-1}\right]^{-1}=
\frac{1}{4}\left( 
\begin{array}{cc}
\hat{t}\hat{t}^\dagger & 0 \\
0 & \hat{t}^\prime\hat{t}^{\prime\dagger} \end{array} 
\right),
\label{eq:rank}
\end{equation}
every small eigenvalue of $\hat{M}\hat{M}^\dagger$ should have the counterpart of $\hat{t}\hat{t}^\dagger$ or $\hat{t}^\prime\hat{t}^{\prime\dagger}$. Likewise, $2 N_{min}$ finite eigenvalues of $\hat{M}\hat{M}^\dagger$ corresponds to $N_{min}$ finite eigenvalues of $\hat{t}\hat{t}^\dagger$ and $N_{min}$ of $\hat{t}^\prime\hat{t}^{\prime\dagger}$. Therefore, if certain eigenvalues become small in some of $\hat{M}_i$ (e.g. in a constriction) they remain small in $\hat{M}$ and in both $\hat{t}\hat{t}^\dagger$ and $\hat{t}^\prime\hat{t}^{\prime\dagger}$. These small eigenvalues correspond to the {\it evanescent eigenchannels}. They are distinct from closed eigenchannels with $\tau>\tau_{C}\sim\exp[-L/\ell]\gg\exp[-d/\lambda]$. The latter inequality follows from $\lambda\ll\ell$.

In systems with absorption, in contrast, optical reciprocity is the only constraint left. It leads to $\hat{t}=\hat{t}^{\prime T}$, $\hat{r}=\hat{r}^{T}$, $\hat{r}^\prime=\hat{r}^{\prime T}$, where superscript $T$ denotes matrix transpose. Despite the fact that non-zero eigenvalues of $\hat{t}^\dagger\hat{t}$ and $\hat{t}^{\prime\dagger}\hat{t}^\prime$ are still identical, they are no longer related to those of the reflection matrices because input energy can be absorbed instead of being transmitted or reflected.

\subsection{Numerical simulations\label{subsec:numerical}}

\begin{figure}[htbp]
\centering
\vskip -0.0cm
\includegraphics[width=2.5in]{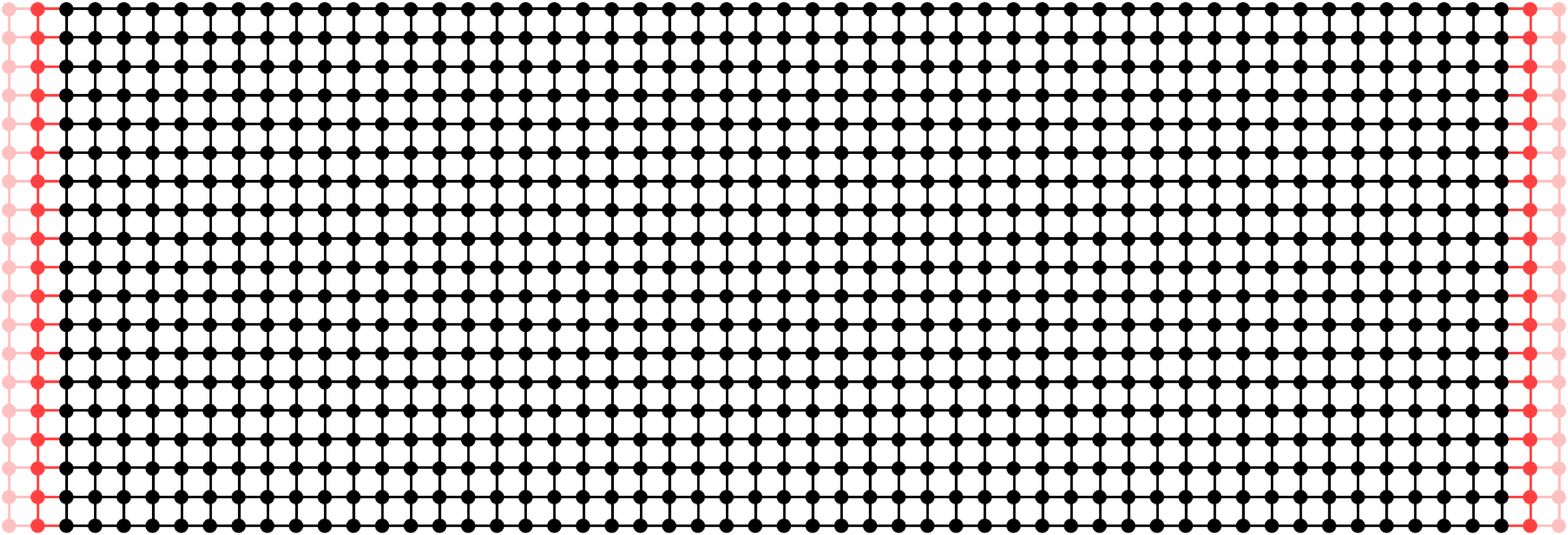}
\vskip -0.1cm
\caption{\label{fig:waveguide} Transport in a waveguide, which consists of a collection of coupled lattice sites, is described by the Hamiltonian in Eq.~(\ref{eq:H}). The disordered region (black dots) is connected to two infinitely long leads (red dots) on the left and right.}
\end{figure}

Numerical simulations are based on Kwant software package~\cite{2014_Groth_Kwant}. It allows one to conveniently compute $\hat{S}$ matrix of disordered waveguide defined as a collection of coupled lattice sites $|i\rangle$ in two dimensional grid described by a tight-binding Hamiltonian
\begin{equation}
\hat{H}|\psi\rangle=\left[\sum\limits_{i,j}H_{ij}|i\rangle\langle j|\right]
\left( \sum\limits_l \psi_l |l\rangle \right),
\label{eq:H}
\end{equation}
where $\psi_l$ is the wavefunction (i.e. field) amplitude at site $l$ (a 2D vector), see Fig.~\ref{fig:waveguide}. We introduce disorder by adding a random on-site potential $\delta E_{ii}$ to the diagonal elements $H_{ii}=E_0+\delta E_{ii}$, while keeping the nearest neighbor coupling at constant value of $1$. The scattering region $0\leq z\leq L$ is connected to the leads -- infinitely long waveguides with $\delta E_{ii}=0$.   

%%%%%%%%%%%%%%%%%%%%%%%%%%%%%%%%%%%%%%%%%%%%%%%%%%%%%%%%%%%%%%%%
%%%%%%%%%%%%%%%%%%%%%%%%%%%%%%%%%%%%%%%%%%%%%%%%%%%%%%%%%%%%%%%%

The model is well suited to describe continuous wave scattering phenomena as long as $k\ell\gg 1$. In our simulation we choose the parameter $E_0$ and the disorder strength $\Delta$ ($-\Delta<\delta E_{ii}<\Delta$) so that $k\ell\simeq 60$. Because $\langle \delta E_{ii}\rangle =0$, the average value of $H_{ii}$ does not change with the strength of disorder, avoiding any impedance mismatch between the leads and the random waveguide~\cite{2013_Tian_Eigenvalues}. To eliminate the ballistic component that propagates through the system without scattering, we set $L/\ell\simeq 30 \gg 1$. 

The transport mean free path $\ell$ is obtained from the value of dimensionless conductance $g_{p}$ statistically averaged over an ensemble of disorder realizations. In passive waveguide of constant width $g_{p}=(\pi/2)N\ell/L$ uniquely determines $\ell$. 

To ensure diffusive transport ($g_p\gg 1$) in the simulated waveguides, we select the number of modes (i.e. the width of the waveguides) to be sufficiently large - on the order of hundreds.

Absorption is introduced to our system by adding a small constant negative imaginary part $\gamma$ to the on-site potential $E_0\rightarrow E_0+i\gamma$. The specific value of $\gamma$ is selected to obtain the desired value of the diffusive absorption length $\xi_a$. The latter is obtained from $g=(\pi N\ell/\xi_a)\exp[-L/\xi_a]$ for a constant width waveguide with $L/\xi_a\gg 1$~\cite{1998_Brouwer}.

To obtain the densities of transmission, reflection and absorption eigenvalues in Figs.~1-4, we perform simulations for 1000 random realizations of $\delta E_{ij}$, giving us $\sim 2\div 5\times 10^6$ eigenvalues with a sufficiently low noise over six decades of magnitude.

\subsection{Density of the absorption eigenvalues with two-sided excitation\label{subsec:PofSabs}}

%%%%%%%%%%%%%%%%%%%%%%%%%%%%%%%%%%%%%%%%%%%%%%%%%%%%%%%%%%%%%%%%
\begin{figure}[htbp]
\centering
\vskip -0.0cm
\includegraphics[width=3.3in]{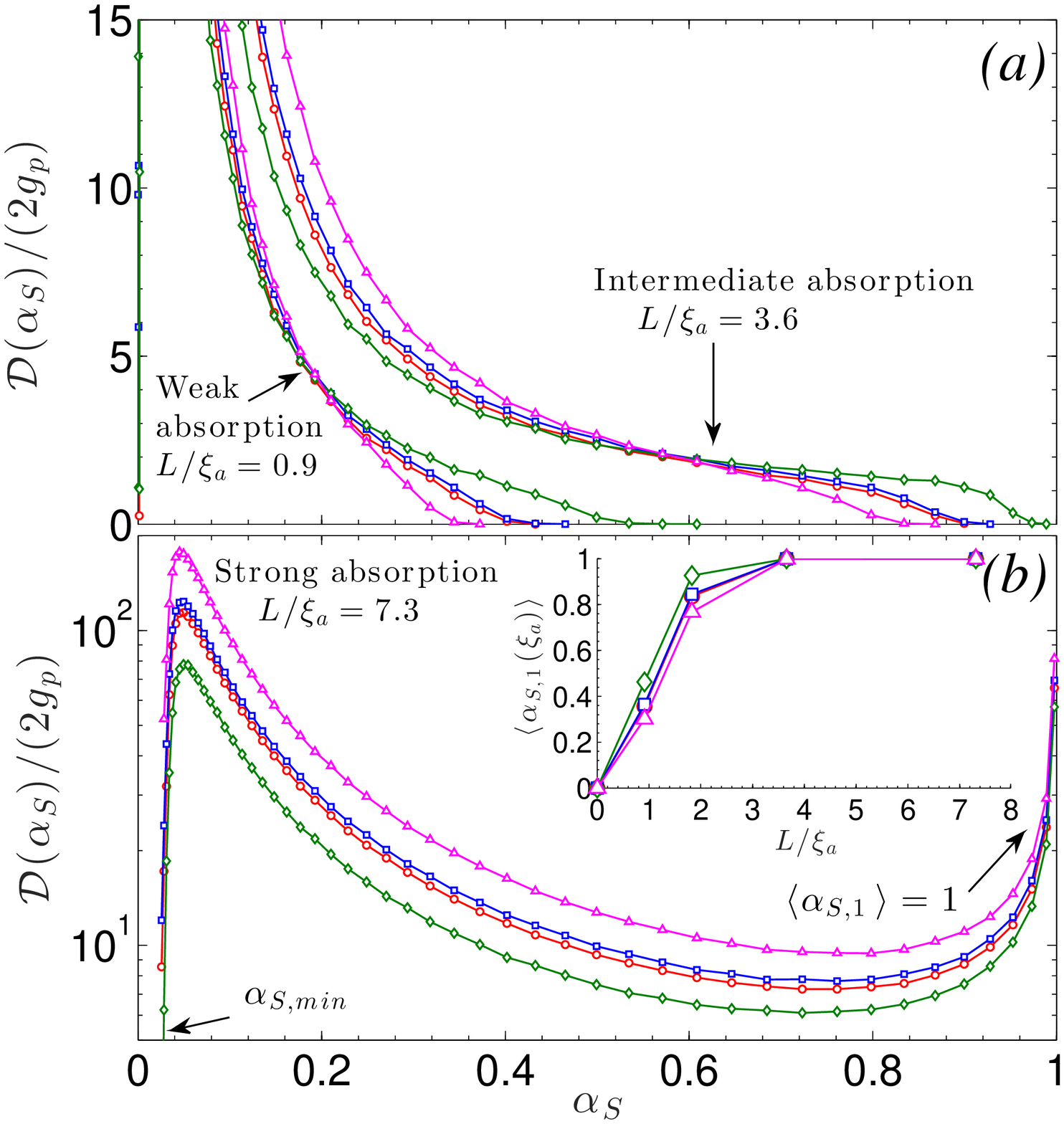}
\vskip -0cm
\caption{\label{fig:PofabsS} 
Density of absorption eigenvalues $\alpha_S$ with two-sided illumination at weak absorption $\langle\alpha_{S,1}\rangle\ll 1$, intermediate absorption $\langle\alpha_{S,1}\rangle\lesssim 1$ (panel a), and strong absorption $\langle\alpha_{S,1}\rangle\simeq 1$ (panel b). Symbol notations for constant width, horn, lantern, and bow-tie waveguides are the same as in Fig.~3 of the main text. With increasing absorption, ${\cal D}(\alpha_S)$ develops a long tail at large $\alpha_S$, which eventually becomes a peak at $\alpha_S \simeq 1$.  In all cases ${\cal D}(\alpha_S)$ shows strong dependence on the shape of the waveguide. The inset of panel (b) plots the ensemble-averaged largest eigenvalue $\langle\alpha_{S,1}\rangle$ vs. the absorption strength $L/\xi_a$.}
\end{figure}
%%%%%%%%%%%%%%%%%%%%%%%%%%%%%%%%%%%%%%%%%%%%%%%%%%%%%%%%%%%%%%%%

Fig.~4 in the main text shows the density of $\alpha$ -- eigenvalues of $\hat{A}\equiv\hat{I}-\hat{t}^\dagger\hat{t}-\hat{r}^\dagger\hat{r}$ for waves incident from the left lead or $\hat{A}^\prime\equiv\hat{I}-\hat{t}^{\prime\dagger}\hat{t}^\prime-\hat{r}^{\prime\dagger}\hat{r}^\prime$ for the wave incident from the right lead. In contrast, the eigenvalues $\alpha_S$ of the matrix $\hat{A}_S\equiv\hat{I}-\hat{S}^\dagger\hat{S}$ correspond to eigenvectors with waves incident from both leads simultaneously -- {\it two-sided excitation}, see Eq.~(\ref{eq:S}). 

Figure~\ref{fig:PofabsS} shows evolution of the density of eigenvalues $\alpha_S$, ${\cal D}(\alpha_S)$, with absorption. While it is qualitatively similar to ${\cal D}(\alpha)$  in Fig.~4 of the main text, we notice the following differences. First, unlike one-sided excitation case where ${\cal D}(\alpha)$ can be dependent of the input direction in an asymmetric waveguide (e.g. ${\cal D}_L(\alpha)\neq {\cal D}_R(\alpha)$ in a horn waveguide), there is only one ${\cal D}(\alpha_S)$. The number of $\alpha_{S,n}$ in any given sample is $N_L+N_R$, greater than that of $\alpha_n$ ($=N_L$) or of $\alpha_n^\prime$ ($=N_R$). Secondly, in the regimes of weak and intermediate absorption, the density of ${\cal D}(\alpha_S)$ exhibits long tails toward $\alpha_S\rightarrow 1$, c.f. Fig.~\ref{fig:PofabsS}a, whereas ${\cal D}(\alpha)$ has already developed a second peak, c.f. Fig.~4a of the main text. Third, because of the tail of ${\cal D}(\alpha_S)$, eigenvalues close to unity ($\alpha_S\simeq 1$) can be found at the amount of absorption smaller than that for ${\cal D}(\alpha)$, which exhibits a sharp dropoff beyond the second peak. The latter conclusion can be understood intuitively: two-sided excitation allows for a greater degree of input control. In strong absorption regime, our results coincide with the results of Ref.~\cite{2011_Stone_Coherent_Absorption}, where the concept of coherent enhancement of absorption was first proposed.

\subsection{Existing theoretical description of ${\cal D}(\tau)$ \\
in constant width waveguides with absorption \label{subsec:Poftau_theory}}

I%%%%%%%%%%%%%%%%%%%%%%%%%%%%%%%%%%%%%%%%%%%%%%%%%%%%%%%%%%%%%%%%
\begin{figure}[htbp]
\centering
\vskip -1cm
\includegraphics[width=3.3in]{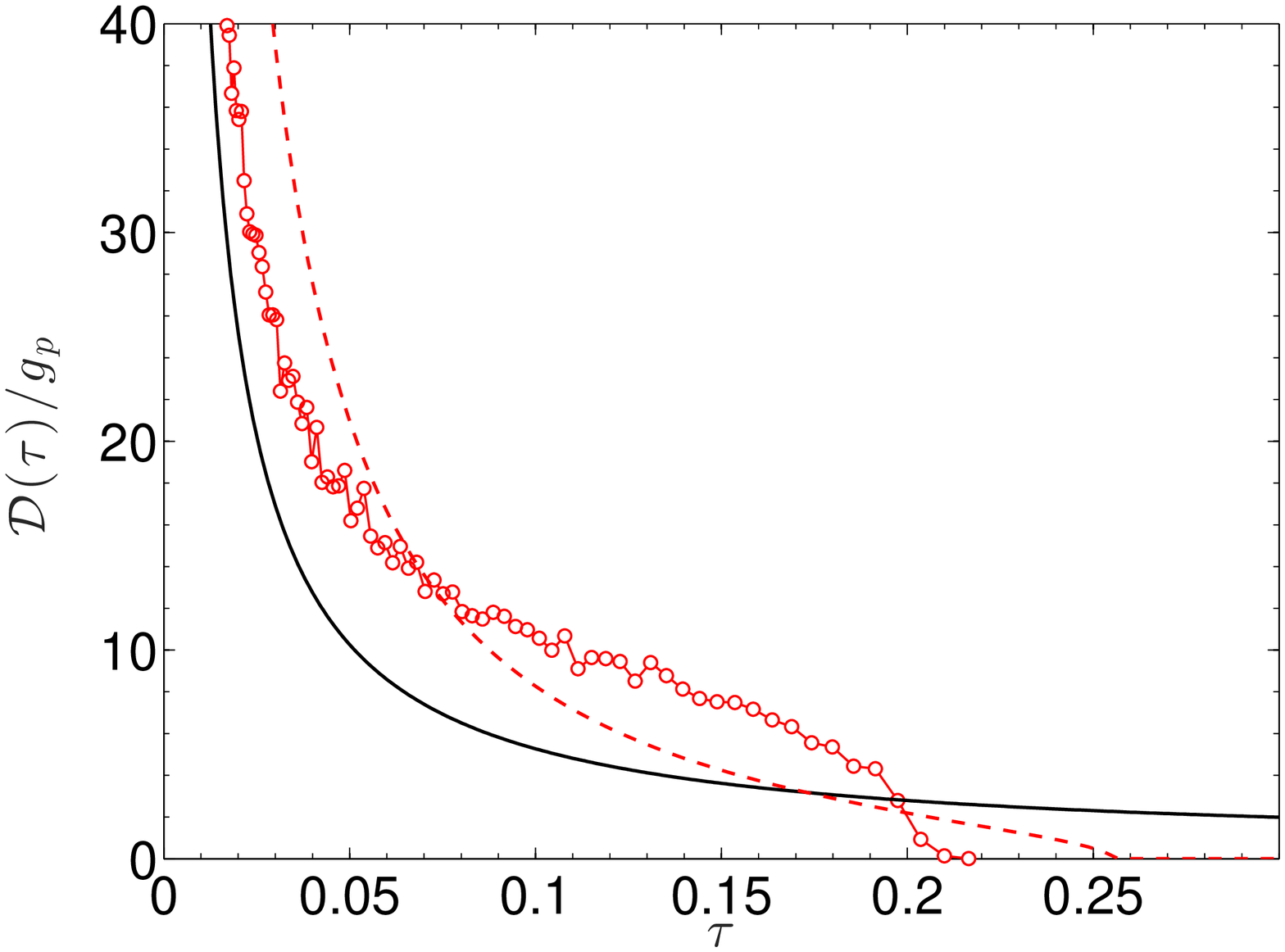}
\vskip -0cm
\caption{\label{fig:Poftau_theory} 
Normalized density of the transmission eigenvalues ${\cal D}(\tau)/g_p$ in constant width waveguide with absorption $L/\xi_a=3.6$ (circles). Solid line is the bimodal distribution shown for reference. Dashed line plots Eqs.~(\ref{eq:Poftau_theory},\ref{eq:Gofz_Brouwer}) (not a fit) for $L/\xi_a=3.6$ sample.}
\end{figure}
%%%%%%%%%%%%%%%%%%%%%%%%%%%%%%%%%%%%%%%%%%%%%%%%%%%%%%%%%%%%%%%%

Fig.~2 of the main text we present the results of numerical simulations of ${\cal D}(\tau)$ in absorbing disordered waveguides. In this section we briefly review the previous theoretical model in Ref.~\cite{1998_Brouwer}, and apply it to describe the density of transmission eigenvalues in {\it constant width waveguides}. 

We begin by introducing a resolvent 
\begin{equation}
G(z)=\left\langle{\rm Tr}\frac{1}{z-\hat{t}^\dagger\hat{t}}\right\rangle, 
\label{eq:resolvent}
\end{equation}
which formally defines the eigenvalue density as
\begin{equation}
{\cal D}(\tau)=-(1/\pi){\rm Im} [G(\tau+i0)].
\label{eq:Poftau_theory}
\end{equation}
In passive systems, $G(z)$ has been found in several works (see Ref.~\cite{2009_Nazarov_Book_Quantum_Transport} for review):
\begin{equation}
G_0(z)=\frac{N}{z}-\frac{g_p}{z\sqrt{1-z}}{\rm arctanh}\left[\frac{{\rm tanh}(N/g_p)}{\sqrt{1-z}}\right],
\label{eq:G0ofz}
\end{equation}
which leads to the bimodal distribution in the main text.

In Ref.~\cite{1998_Brouwer}, Brouwer obtained an approximate analytical expression for $G(z)$, and hence for ${\cal D}(\tau)$ in the limit of strong absorption $L/\xi_a\gg 1$. The result reads
\begin{equation}
G^{(B)}(z)=\frac{N}{z}-\frac{2g_p}{z}\ {\cal W}\left(-\frac{L}{\xi_a z}e^{-L/\xi_a}\right),
\label{eq:Gofz_Brouwer}
\end{equation}
where ${\cal W}(z)$ is Lambert W-function. This expression is plotted in Fig.~\ref{fig:Poftau_theory} with a dashed line for $L/\xi_a=3.6$. It is not a fit, because all parameters in Eq.~(\ref{eq:Gofz_Brouwer}) are known independently. We observe that Eqs.~(\ref{eq:Poftau_theory},\ref{eq:Gofz_Brouwer}) overestimates the maximum value of $\tau$ and underestimates the density at small $\tau$ ($\lesssim 0.2$). We attribute these deviations to insufficiently large value of the absorption parameter $L/\xi_a$, which does not reach the strong absorption limit required for deriving  Eq.~(\ref{eq:Gofz_Brouwer}). Lastly, it is not immediately clear how this model can be generalized to the waveguides with varying width.

\bibliography{../../Bibliography/latex_bibliography}
%\bibliography{../../../Bibliography/latex_bibliography}
%\bibliography{2015_Eigenchannels_SI0.bbl}